# Origin and enhancement of the 1.3 µm luminescence from GaAs treated by ion-implantation and flash lamp annealing

Kun Gao[1,2], S. Prucnal[1], W. Skorupa[1], M. Helm[1,2], Shengqiang Zhou[1]

1. Institute of Ion Beam Physics and Materials Research, Helmholtz-Zentrum Dresden-Rossendorf (HZDR), P.O. Box 510119, 01314 Dresden, Germany
2. Technische Universität Dresden, 01062 Dresden, Germany

**Abstract**

GaAs and GaAs based materials have outstanding optoelectronic properties and are widely used as light emitting media in devices. Many approaches have been applied to GaAs to generate luminescence at 0.88, 1.30, 1.55 µm which are transmission windows of optical fibers. In this paper we present the photoluminescence at 1.30 µm from deep level defects in GaAs treated by ion-implantation and flash lamp annealing (FLA). Such emission, which exhibits superior temperature stability, can be obtained from FLA treated virgin GaAs as well as doped GaAs. Indium-doping in GaAs can greatly enhance the luminescence. By photoluminescence, Raman measurements, and positron annihilation spectroscopy, we conclude that the origin of the 1.30 µm emission is from transitions between the $V_{As}$-donor and X-acceptor pairs.

**Introduction**

GaAs and GaAs based materials are widely used in optoelectronic and photovoltaic devices for their direct bandgap as well as outstanding optical and electrical performances. At room temperature, GaAs has a bandgap of 1.42 eV which corresponds to 0.87 µm luminescence [1]. Many approaches have been designed on GaAs related matrices to achieve 1.30 µm and 1.55 µm emissions which are in the 2$^{nd}$ and 3$^{rd}$ transmission window of optical fibers. To achieve emissions at such wavelengths, one way is to modify the bandgap by alloying, *i.e.* to form ternary/quaternary alloys such as InGaAs/InGaAsP [2]. Another common method is to grow multiple quantum wells or quantum dots with GaAs and some other narrower bandgap materials (*e.g.*, InAs), using quantum confinement to control the emission wavelength [3,4]. For practical applications, the stability at a broad temperature range is one of the key indicators for the device performance. Besides these two approaches, proper defect engineering can also be used for the photoluminescence (PL) enhancement below the band gap of GaAs. The defect related luminescence bands in GaAs are in the spectral range of 0.8 – 1.35 eV, consisting of gallium and/or arsenic vacancy/interstitial complexes [5,6]. Related to the nature of a defect level, its luminescence can also be relatively immune to temperature changes. The origin of most defect related emission bands in GaAs is quite well established in literature except the 0.95 eV (1.30 µm) emission whose origin is still controversial. Considering the fact that light at a wavelength of 1.30 µm traveling through quartz fiber suffers minimum attenuation, such a GaAs based light source could be applied in the field of optical fiber communications. For industrial production, the cost and the efficiency are two decisive factors. Ion-implantation and post-implantation thermal annealing (*e.g.*, rapid thermal annealing (RTA) and flash lamp annealing (FLA)) [7,8] are compatible techniques in IC-industry. Therefore, to realize the 1.30 µm emission from GaAs-based materials by ion-implantation and FLA is promising for its advantages of low-cost, high efficiency, and IC-compatibility.

In previous work [9] we have presented defect engineering in nitrogen-doped and undoped semi-insulating GaAs (SI-GaAs) wafers by millisecond range FLA for efficient room temperature 1.30 µm PL. According to the experimental results, the 1.30 µm PL observed from a FLA treated virgin GaAs wafer is due to radiative transitions between energy levels of intrinsic defects in GaAs, and this emission can be enhanced by nitrogen implantation followed by FLA [9]. In this paper we present a detailed study of the influence of N, P, In, and Zn implantation combined with FLA on the 1.30 µm PL emission from GaAs. Different doping types are used for comparison to investigate the origin of the 1.30 µm PL emission. Proper N-, P-, and In-doping leads to the formation of a GaAs based ternary alloy, among which N and P tend to occupy the As site while In goes to the Ga site, which will result in the significant differences on defect-type in GaAs. Zn-doping leads to the conduction type change in GaAs (*i.e.*, from semi-insulating to p-type conducting). Our results show that indium-doping has greatly enhanced the emission at 1.30 µm by more than two orders of magnitude compared with FLA treated virgin SI-GaAs. Moreover, the influence of the conductivity type on the luminescent properties of the GaAs wafers is also discussed. In the case of Zn-doping, the 1.30 µm emission is completely quenched, which is consistent with our previous results shown for p-type Mn-doping [9].

**Experimental Setup**

Semi-insulating (100) GaAs wafers were implanted at room temperature with N, P, In, and Zn ions. The dopants are implanted with different kinetic energies to ensure that the concentration depth profiles of the dopants are almost the same. Nitrogen ions are implanted deeper considering the diffusion of nitrogen during annealing. The as-implanted and virgin GaAs wafers were annealed by a flash lamp system at different energy densities (*i.e.*, different annealing temperatures on the sample surface) for 3 ms or 20 ms. The annealing energy density was controlled by changing the power supplied to the Xe lamps used during FLA. For the whole sample series, the optimized annealing condition for the 1.30 µm PL emission presented in this paper are slightly below the melting point of the virgin/as-implanted GaAs. The implantation and annealing parameters are given in Table 1. To prevent the decomposition of the GaAs surface and the evaporation of As during annealing, 200 nm thick $SiO_2$ layers were deposited on the surface of the GaAs wafers by PECVD at 200 °C before annealing, and then were chemically etched in $HF:H_2O$ solution after annealing.

Optical properties of the virgin and annealed GaAs samples were investigated by temperature dependent PL. Micro-Raman spectroscopy was used to determine the structural properties of GaAs and the influence of doping before and after FLA. The PL measurements were performed using a 532 nm Nd:YAG laser with an intensity of about 3 W/cm$^2$ for sample excitation at temperatures from 20 K to 300 K. The PL signal was dispersed by a Jobin Yvon Triax 550 monochromator and recorded by a liquid-nitrogen-cooled InGaAs detector. The Raman spectra were collected in a backscattering geometry in the range of 150 to 600 cm$^{-1}$ by a liquid-nitrogen-cooled charge coupled device with 532 nm Nd:YAG laser excitation.

**Table I** Sample preparation: fluence and kinetic energy for different ion-implantations, flash lamp annealing time and energy densities for post-treatment.

| Dopants | | - (virgin) | N* | | P | In | Zn |
|---|---|---|---|---|---|---|---|
| Ion-implantation | Fluence (cm$^{-2}$) | - | $2\times10^{15}$ | $8\times10^{15}$ | $2\times10^{15}$ | $2\times10^{15}$ | $2\times10^{15}$ |
| | Energy (keV) | - | 30 | 70 | 40 | 140 | 120 |
| | Doping range (nm) | - | 250 | | 90 | 90 | 100 |
| | Peak concentration (at. %) | - | 2 | | 1.6 | 1.8 | 1.4 |
| Flash lamp annealing | Time (ms) | 20 | 3 | | 20 | 20 | 20 |
| | Optimized energy density (J/cm$^2$) | 89 | 53 | | 89 | 89 | 89 |

\* Nitrogen was implanted deeper than other dopants due to its diffusion and evaporation during FLA. A double-implantation (*i.e.,* dopants are implanted with two different kinetic energies) was applied in order to form a thick and homogenous doping layer.

**Results and Discussion**

**a. Photoluminescence**

**(1) Virgin GaAs**

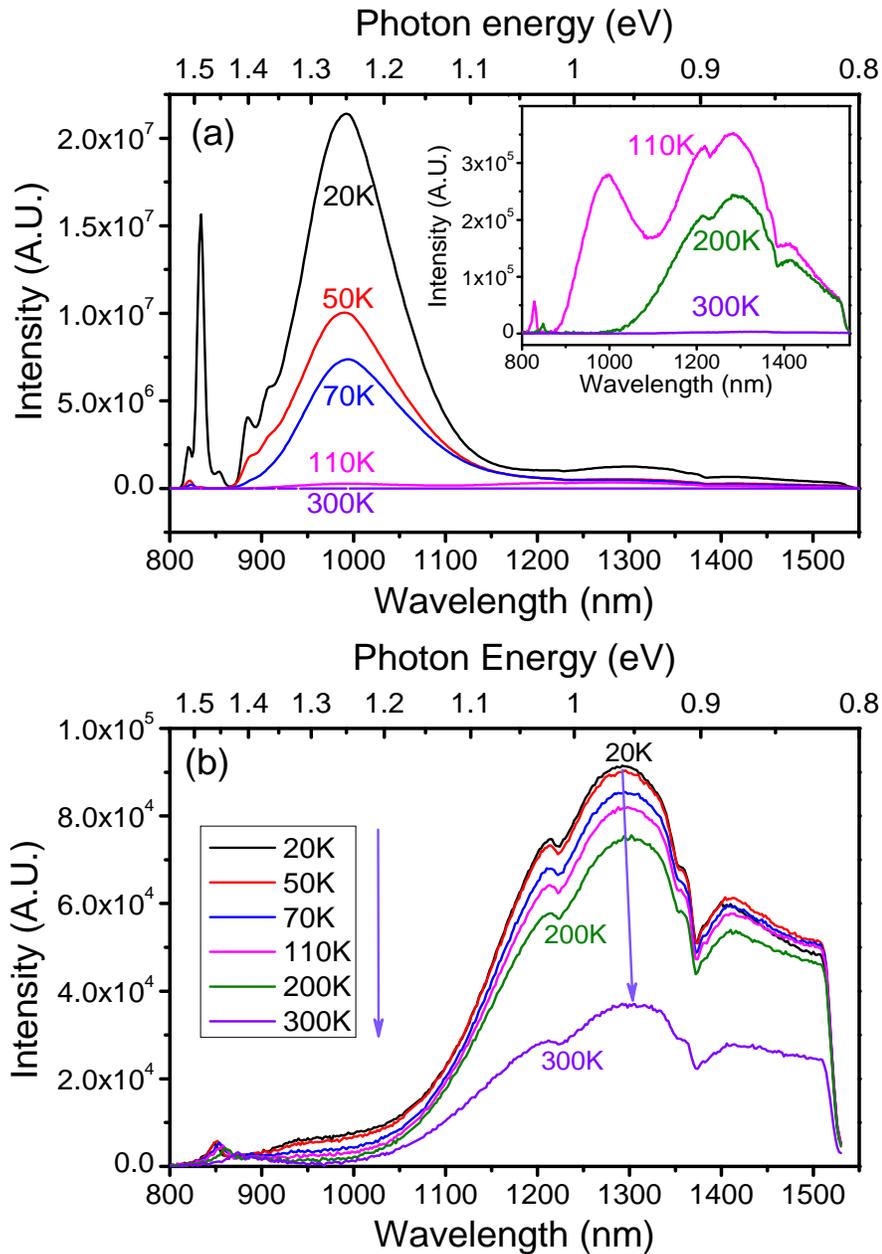

Fig. 1 PL spectra of virgin (a) and 20 ms flash lamp annealed (energy density 89 J/cm$^2$) (b) SI-GaAs wafer for different temperatures as indicated. Inset of (a) shows the magnified PL spectra of virgin GaAs at 110 K, 200 K, and 300 K. The cut-off at around 1520 nm is due to the limit of the InGaAs detector.

Figure 1(a) shows the temperature dependent PL spectra obtained from a virgin SI-GaAs wafer measured between 20 K to 300 K. The PL spectra are composed of several strongly temperature dependent peaks related to the near band edge emission (NBE) and defect centers in GaAs. The luminescence peak at 820 nm (at 20 K) shows a significant redshift and intensity decrease with increasing temperature, which is the typical behavior for the NBE. The peaks at 834 nm and 992 nm quench faster than the NBE peak, but no significant redshift is observed. The two PL peaks are assigned to transitions from the conduction band to energy levels of the carbon acceptor ($C_{As}$) and Ga vacancy ($V_{Ga}$), respectively [10-14]. Moreover, at 20 K a broad peak at about 1.30 µm is observed. The trough at about 1.38 µm that occurs in each spectrum is attributed to the absorption of water existing in the PL system. The other small trough at about 1.23 µm is unlikely due to the absorption of water since it is absent when measuring other materials which have PL in this range (e.g., boron doped silicon, not shown) by the same PL system. A distinct peak at around 1.22 µm appears as an accompanying peak of the 1.30 µm peak. Further discussions of the two peaks will be presented in section c. The intensities of both the 1.30 µm and the 1.22 µm peaks decrease with rising temperature. At 300 K, the two peaks almost vanish due to the thermal quenching, *i.e.* the intensity is decreased by three orders of magnitude.

After flash lamp annealing, the PL spectrum (see Fig. 1 b) differs significantly from the spectrum of the non-annealed sample (Fig. 1 a). Except for the two NIR emissions at around 1.30 µm and 1.22 µm, most of the PL peaks from the non-annealed sample either disappear or become much weaker compared with the virgin wafer. In contrast to the complete thermal quenching of the $C_{As}$ and $V_{Ga}$ defect-related luminescence (834 nm and 992 nm, respectively), both NIR peaks at 1.22 and 1.30 µm persist at 300 K. Though the two peaks still maintain a decay trend as temperature increases, they show only 60% intensity decrease from 20 K to 300 K, which is remarkably different from the virgin sample before annealing. At the same time, the two NIR peaks show only 14 nm redshift.

## (2) Ion-implanted GaAs

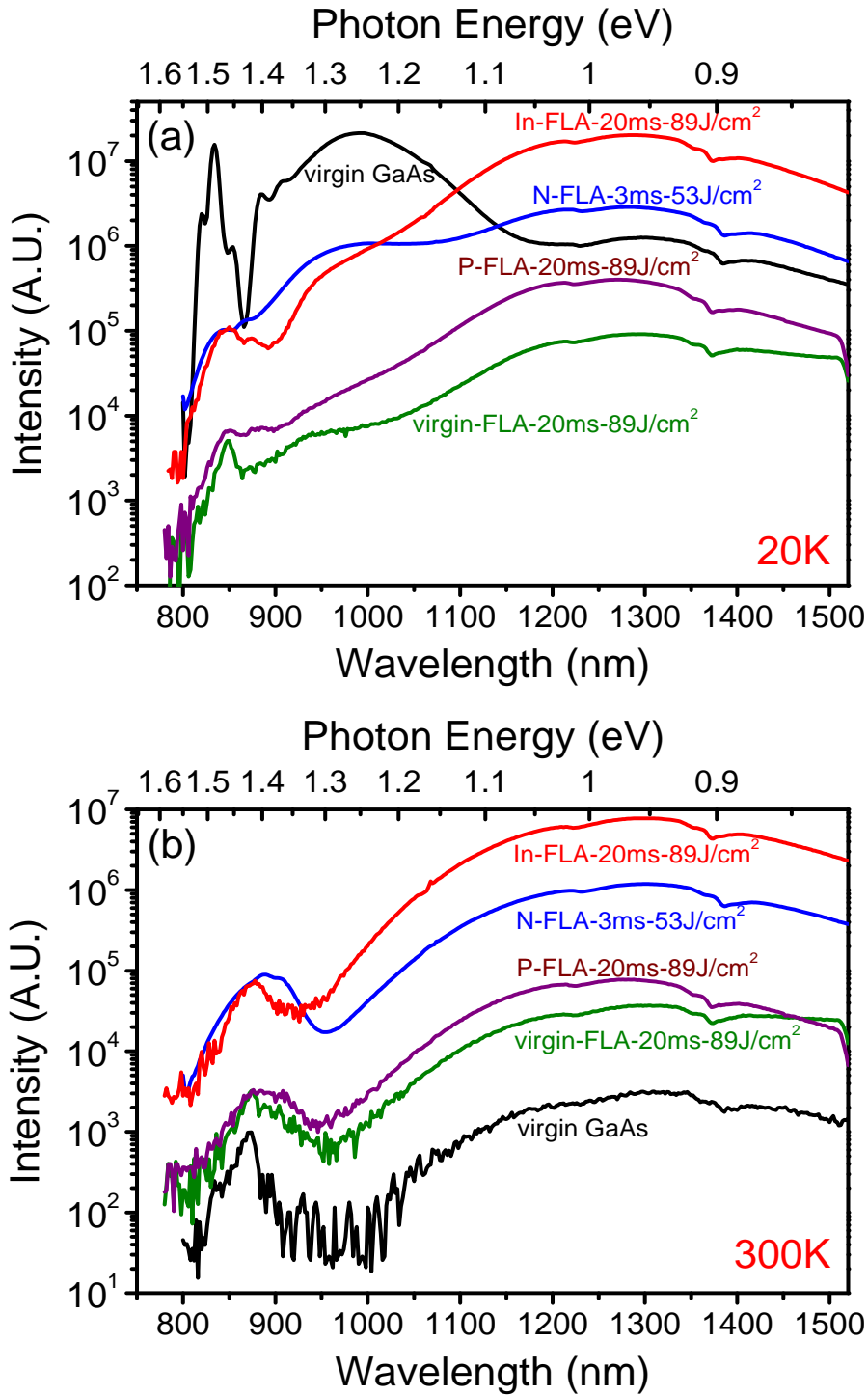

Fig. 2 PL spectra measured at 20 K (a) and 300 K (b) from In-, P-, N- implanted GaAs treated by FLA. PL from the non-annealed virgin GaAs as well as the FLA treated virgin

GaAs at 20 K and 300 K (plotted on linear scale in Fig. 1 (a,b)) is shown for comparison. All spectra show the 1.30 µm emission accompanied by the 1.22 µm emission with a similar line shape but different intensities.

Figure 2 shows PL spectra from FLA-treated virgin, N-, P-, In- implanted GaAs as well as from a virgin GaAs wafer tested at 20 K (a) and 300 K (b). For the spectra at 300 K the PL line shapes for different samples are similar: the 1.30 µm peak accompanied by the 1.22 µm peak dominates, while a weaker NBE appears at around 875 nm.

For FLA treated nitrogen-implanted GaAs (marked as N-FLA-3ms-53J/cm$^2$) at 20 K, the PL peak at 842 nm weakens and redshifts as the temperature increases (not shown here), which can be attributed to the NBE from GaAs. The broad peak at about 0.99 µm can be attributed to the transition between conduction band to Ga-vacancy ($V_{Ga}$), as shown and discussed for the virgin GaAs sample. For the 1.30 µm emission, the behavior is similar to the flash lamp annealed virgin GaAs shown in Fig. 1(b), *i.e.*, 60 % intensity reduction and 8 nm redshift from 20 K to 300 K.

At around 850 nm one PL peak is also detected at 20 K from FLA treated phosphorus-implanted GaAs (marked as P-FLA-20ms-89J/cm$^2$), corresponding to the NBE from GaAs. The emission from $V_{Ga}$ for P-implanted GaAs is rather weak at all temperatures, which is the main difference from the annealed GaAs:N samples. The 1.30 µm emission shows stronger quenching in comparison to the virgin and N implanted sample. Between 20 K and 300 K the intensity decreases by 80% and shifts by 12 nm.

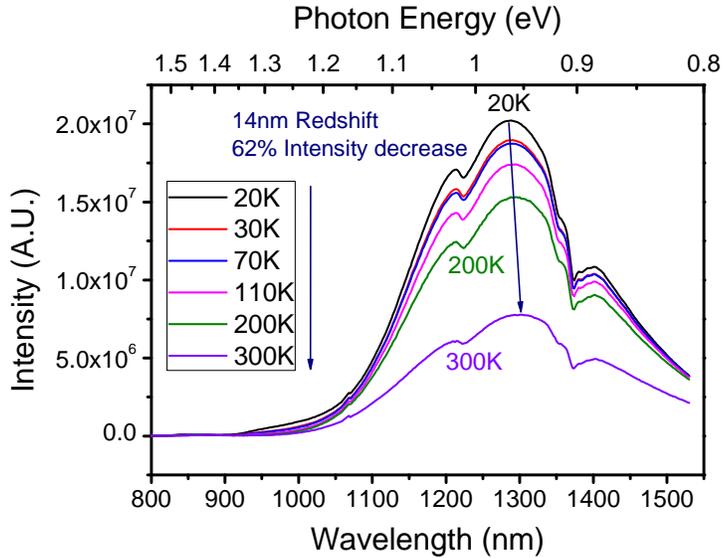

Fig. 3 PL spectra of indium-doped GaAs treated by flash lamp annealing for 20 ms at 89 J/cm$^2$. The 1.30 μm and 1.22 μm peaks dominate between 20 K and 300 K.

For the FLA treated GaAs:In sample (marked as In-FLA-20ms-89J/cm$^2$), the 1.30 μm emission is greatly enhanced (see Fig. 2 and Fig. 3). The alloying of 1.8 at. % of indium with GaAs can bring about 26 meV bandgap shrinkage, which corresponds to 17 nm redshift of the NBE PL [15]. Therefore the 1.30 μm emission cannot originate from transitions between the conduction and valence band of InGaAs. The main difference between the incorporation of indium and group V elements into GaAs is that indium replaces gallium in the matrix, while group V elements substitute arsenic. The substitution of indium has been confirmed by Rutherford backscattering channeling spectrometry (not shown). Therefore indium-doped GaAs differs from nitrogen- and phosphorus-doped GaAs in the type and the concentration of the defects. At 20 K we observed a very weak NBE peak at about 840 nm with the typical redshift for the NBE emission as the temperature increases. In the spectra only the 1.30 μm emission accompanied by the 1.22 μm peak dominates. The $V_{Ga}$ peak at about 0.99 μm appears as a shoulder. From 20 K to 300 K a 14 nm redshift and 60 % intensity decrease occur, which suggests good thermal stability of the 1.30 µm luminescence.

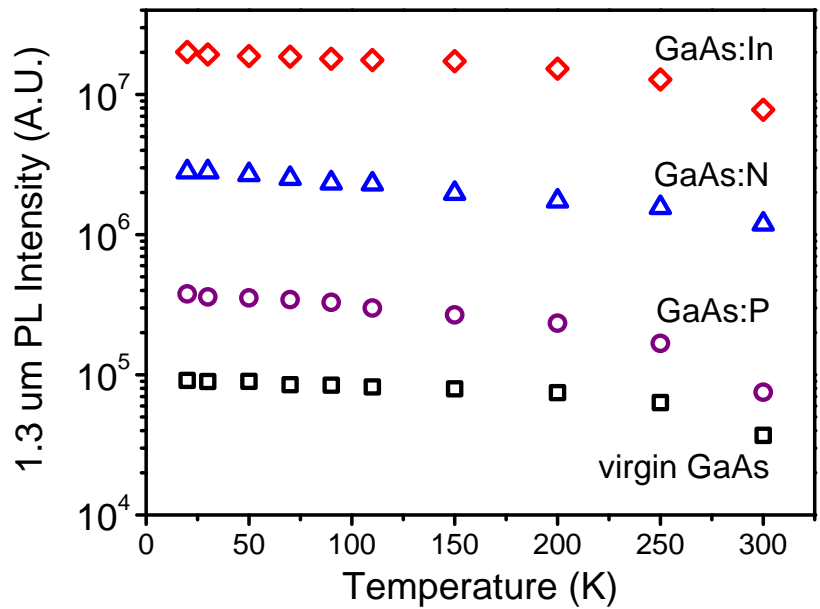

Fig. 4 Intensities of the 1.30 µm PL emission from FLA-treated GaAs:In, GaAs:P, GaAs:N as well as virgin GaAs as the function of temperature. For a better view the vertical axis is in logarithmic scale.

Figure 4 shows the intensity differences and the variation tendencies with increasing temperature of 1.30 µm emissions from different samples. As the temperature increases, all the samples show monotonic decrease in the emission intensity. The annealed virgin GaAs has the weakest 1.30 µm luminescence while the group-V implanted GaAs samples show an enhancement of this emission. Indium-doped GaAs exhibits the strongest 1.30 µm emission which is more than two orders of magnitude higher than that observed from the virgin GaAs. Compared with the thermal quenching of the 1.30 µm PL from the non-annealed virgin GaAs wafer (see Fig. 1a), the FLA treated samples only show a limited intensity reduction. Therefore we conclude that FLA can induce the 1.30 µm luminescence at room temperature in virgin and implanted GaAs samples, *i.e.,* to stabilize such emission from temperature influence.

It is also worth to note that for the virgin GaAs and GaAs implanted by P and In, a stronger 1.30 µm PL is observed from 20 ms FLA treated samples than from 3 ms FLA treated ones, whereas for N-implanted GaAs 3 ms annealing time is superior to 20 ms (not shown here). This is probably due to the instability of gaseous N and its low solubility in GaAs. During high-temperature FLA, N is easy to evaporate from the as-implanted wafer due to its low solubility and high combination energy with gallium in GaAs. Therefore shorter annealing times introduce less negative effects for N-implanted GaAs.

## b. Raman

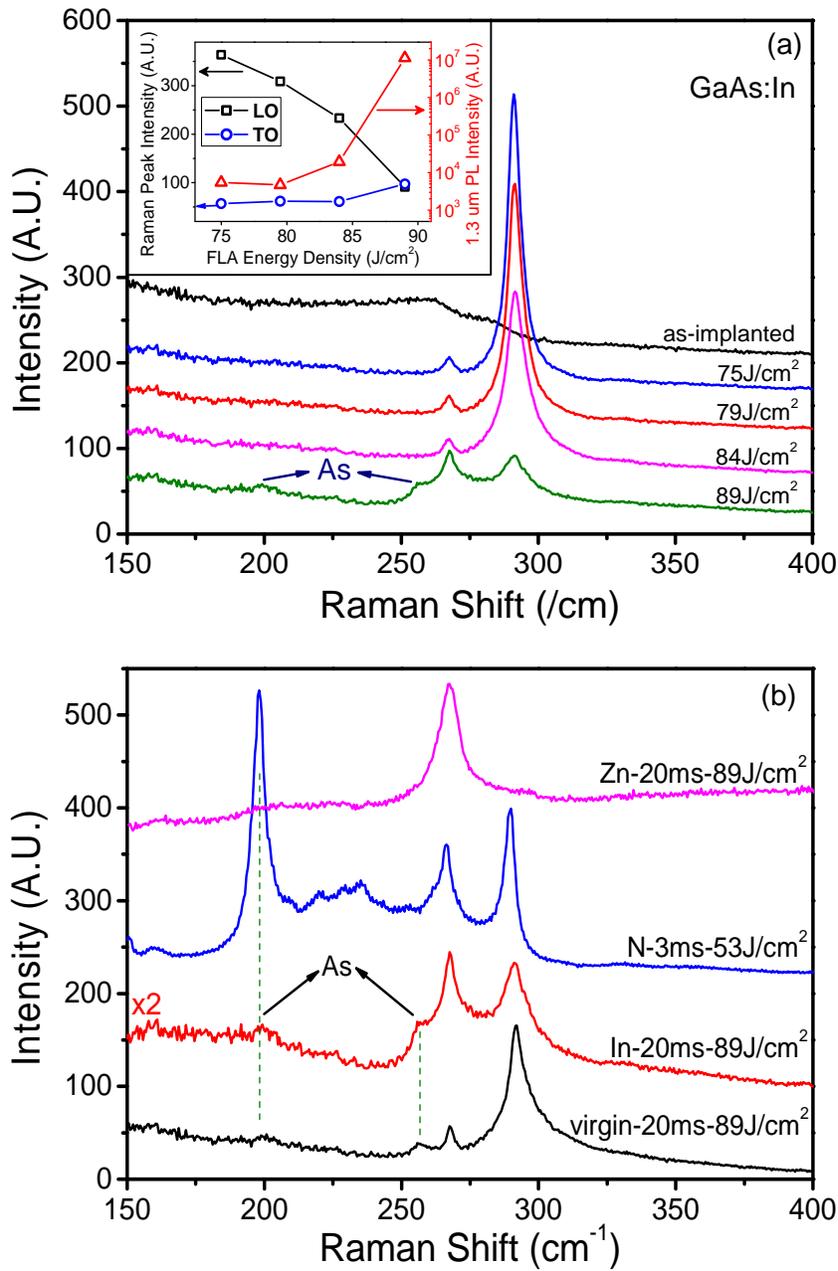

Fig. 5 (a) Raman spectra from indium-implanted GaAs, both as-implanted and 20 ms flash lamp annealed at different energy densities are shown. The inset of (a) compares the intensity variations of the 1.3 μm PL and the LO and TO Raman modes with respect to different FLA energy densities. (b) Raman spectra of FLA treated virgin GaAs,

GaAs:In, GaAs:N, GaAs:Zn, under which annealing condition the strongest 1.30 µm PL (for GaAs:Zn the strongest NBE) is observed. The spectra are vertically shifted for better visibility.

The influence of doping and millisecond flash lamp annealing on the microstructure of GaAs was investigated by means of micro-Raman spectroscopy. Fig. 5(a) shows the first-order micro-Raman spectra obtained from non-annealed and flash lamp annealed Indium-implanted GaAs. According to the selection rules in the backscattering geometry from the (100) oriented monocrystalline GaAs, the Raman spectra should reveal only the longitudinal optical (LO) phonon mode at 292 cm$^{-1}$, whereas the transverse optical (TO) phonon mode located at 268.6 cm$^{-1}$ is forbidden.

The Raman spectrum obtained from the indium-as-implanted sample shows two broad peaks at 284 and 258 cm$^{-1}$ corresponding to the LO and TO phonon modes in amorphous GaAs, respectively, due to the destruction of the top layer of the GaAs wafer during ion-implantation. After FLA treatment, such two peaks shifted back to 291 and 267 cm$^{-1}$, close to the standard value of crystalline GaAs, which indicates the FLA-induced regrowth of the lattice. The appearance of the TO mode suggests that the annealed samples are not monocrystalline and contain defects. As the inset of Fig. 5(a) shows, the intensity of LO phonon mode decreases with rising annealing temperature, whereas the TO phonon mode becomes stronger. At the same time, the 1.30 µm PL intensity shows an upward-trend. By changing the annealing energy density from 84 J/cm$^2$ to 89 J/cm$^2$, the LO mode weakens and the TO mode strengthens to a great extent. Simultaneously, the intensity of 1.30 µm PL rises by three orders of magnitude. For the 89 J/cm$^2$ annealed sample, two weak peaks appear at around 255 and 200 cm$^{-1}$, which can be attributed to crystalline arsenic clusters [16]. This suggests that decomposition has occurred to some extent in the near surface region at higher annealing temperature.

Figure 5(b) compares the Raman spectra from the samples exhibiting the 1.30 μm PL emission with optimized annealing condition. The Raman spectrum from Zn-doped GaAs, where the 1.30 μm PL quenches distinctly (see Fig. 6), is shown for comparison. For the virgin GaAs annealed at 89 J/cm$^2$, the differences to the data obtained for 80 J/cm$^2$ shown in the previous work [9] are the appearance of the TO peak and two arsenic peaks. For N-doped GaAs, a strong peak appears at the position of 200 cm$^{-1}$, which is reported to be the longitudinal acoustic (LA) mode derived from N-induced alloy disorder [17]. Compared with the standard wavenumber of the LO mode for bulk GaAs, the peaks revealed a -2 cm$^{-1}$ shift due to lattice shrinkage by N-doping [17,18]. For the Zn-doped p-type GaAs the LO phonon mode of GaAs is almost invisible, while the strongest peak can be attributed to the coupled-LO-phonon-plasmon mode (CLOPM) which is usually optically active in heavily doped p-type semiconductors [19]. According to the influence of hole concentration on the CLOPM peak shift [20], the hole concentration in our Zn-doped GaAs is of the order of 10$^{19}$ cm$^{-3}$.

### c. Origin of the 1.30 μm emission

In our previous paper [9], we have reviewed some discussions on the 1.30 μm (0.95 eV) emission from the literature. The temperature dependence of the 1.30 μm emission is the main difference between those results in the literature and ours. In our case, we observed the 1.30 μm emission from the virgin GaAs and GaAs implanted with various ions after FLA. Therefore, it is reasonable to conclude that the emission is due to defect centers in GaAs rather than the dopants. The excitation intensity dependence of the NBE and 1.30 µm PL spectra have been tested from the FLA treated GaAs:N at 20 K (not shown). The NBE intensity exhibits a linear dependence on the excitation intensity, whereas the 1.30 µm PL intensity can be fitted to be proportional to the square root of the excitation intensity. This square root dependence of the 1.30 µm emission is the typical behavior of the defect-related PL [21].

Moreover, we have observed a great enhancement of the 1.30 µm PL from indium-doped GaAs as compared to group-V doped GaAs and virgin GaAs. Note that indium-implantation increases the total amount of the group-III atoms and make the as-implanted layer relatively group-III-rich compared to the virgin and Group-V implanted GaAs. Therefore, more group-V vacancies (*i.e.* arsenic vacancies ($V_{As}$) in this case) are expected from indium-doped GaAs during the lattice reformation process induced by FLA, whereas the incorporation of group-V elements will not result in such an effect. FLA treated nitrogen and phosphorus doped GaAs samples exhibit stronger 1.30 µm emission than FLA treated virgin GaAs, this is probably due to the increase of $V_{As}$ densities induced by implantation. In addition, for all GaAs samples implanted with different dopant species, the highest 1.30 µm emissions always appear at the samples treated at the highest annealing energy density (*i.e.* approaching the melting point) in each temperature series. The crystalline arsenic peaks can only be found from the corresponding Raman spectra of these samples. Taking all these into consideration, the 1.30 µm emission should be closely related with $V_{As}$.

For further investigations we implanted Zn into the SI-GaAs wafer, and then the sample was treated by FLA for 20 ms with the energy density of 89 J/cm$^2$. In addition, N was also doped into a commercial p-type (Zn-doped) GaAs wafer under the same implantation and FLA conditions for N-doping in SI-GaAs listed in Table 1. The PL spectra of the two samples are shown in Fig. 6, marked as GaAs:Zn, p-GaAs:N, respectively. Zn-implantation leads to heavy p-type doping, which is totally different from the group III or V doping. The results show that such heavy p-type doping completely quenches the 1.30 μm PL. The absence of the 1.30 μm emission at a p-GaAs:N sample also corroborates this conclusion. From these facts we conclude that p-type doping has negative influence on the 1.30 μm PL. Therefore the defects related to 1.30 μm PL should be n-type or neutral.

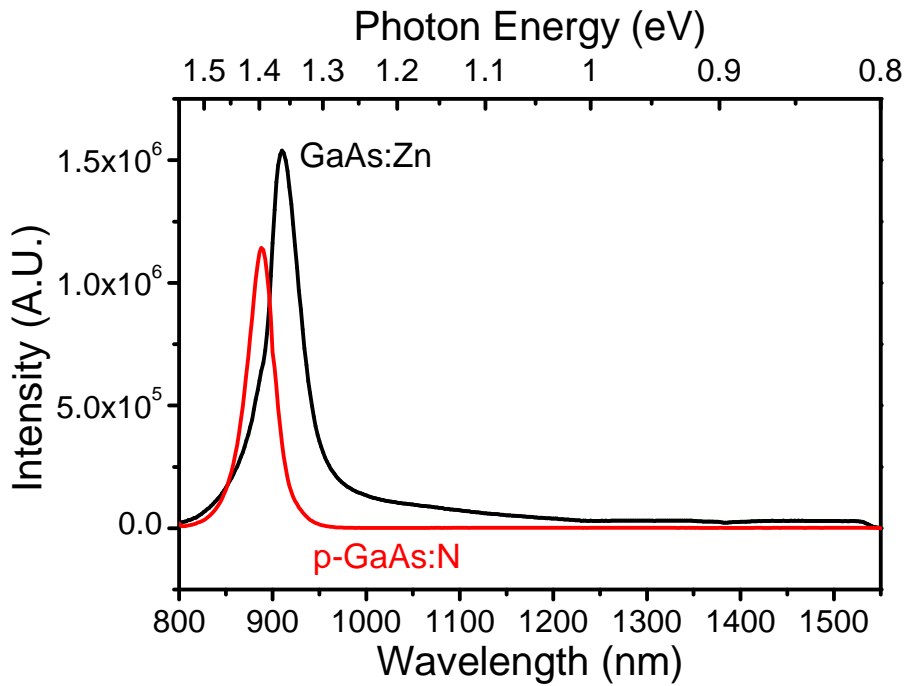

Fig. 6 Room temperature PL spectra of FLA treated GaAs:Zn and p-GaAs:N. The 1.30 μm emission is absent from both samples.

We also investigated the processed GaAs samples which exhibit 1.30 μm PL by positron annihilation spectroscopy (PAS). Results indicate the existence of $V_{As}$-X defects in those samples which exhibit 1.30 μm PL [9]. Considering the fact that the defects detectable by PAS have to be either neutral or negatively charged, while the $V_{As}$ are always positively charged, p-type doping can suppress the recombination of carriers between levels of $V_{As}$ and X by positively charging the X-centers. Note that the arsenic vacancies form shallow donor levels located at about 30, 60, or 140 meV below the conduction band and they are positively charged ($V_{As}^{n+}$) while the X defects' level is located at about 0.5 eV above the valence band [9], the transition between the $V_{As}$-donor and X-acceptor pairs can give rise to the 1.30 µm (0.954 eV) emission. The recombination of electrons and holes between the conduction band and X-acceptor level generates the 1.22 µm emission which always accompanies the main 1.30 µm emission in our samples.

**Conclusions**

In summary, we have presented temperature stable 1.30 µm PL emission from FLA treated undoped and doped GaAs. The highest intensity of such emission was obtained from indium-doped GaAs and is more than two orders of magnitude higher than that from the virgin sample. The origin of this temperature stable 1.30 µm emission is related to a transition between $V_{As}$-X defect complexes. Being generated from defect levels affords the 1.30 µm emission outstanding thermal stability with respect to the emitted wavelength and intensity. Considering the fact that ion-implantation followed by flash lamp annealing is a very convenient and efficient method in industry of chip fabrication and can be easily applied to large scale production, FLA treated GaAs can be a promising candidate for optical-fiber communication devices, especially for those applied in some extreme conditions.

**Acknowledgement**

The authors would like to acknowledge the ion implanter group at Helmholtz-Zentrum Dresden-Rossendorf for ion implantation, and Wolfgang Anwand for the analysis of PAS. The work was financially supported by the Helmholtz-Gemeinschaft Deutscher Forschungszentren (HGF-VH-NG-713).

Fig. 1 PL spectra of virgin (a) and 20 ms flash lamp annealed (energy density 89 J/cm$^2$) (b) SI-GaAs wafer for different temperatures as indicated. Inset of (a) shows the magnified PL spectra of virgin GaAs at 110 K, 200 K, and 300 K. The cut-off at around 1520 nm is due to the limit of the InGaAs detector.

Fig. 2 PL spectra measured at 20 K (a) and 300 K (b) from In-, P-, N- implanted GaAs treated by FLA. PL from the non-annealed virgin GaAs as well as the FLA treated virgin GaAs at 20 K and 300 K (plotted on linear scale in Fig. 1 (a,b)) is shown for comparison. All spectra show the 1.30 µm emission accompanied by the 1.22 µm emission with a similar line shape but different intensities.

Fig. 3 PL spectra of indium-doped GaAs treated by flash lamp annealing for 20 ms at 89 J/cm$^2$. The 1.30 µm and 1.22 µm peaks dominate between 20 K and 300 K.

Fig. 4 Intensities of the 1.30 µm PL emission from FLA-treated GaAs:In, GaAs:P, GaAs:N as well as virgin GaAs as the function of temperature. For a better view the vertical axis is in logarithmic scale.

Fig. 5 (a) Raman spectra from indium-implanted GaAs, both as-implanted and 20 ms flash lamp annealed at different energy densities are shown. The inset of (a) compares the intensity variations of the 1.3 µm PL and the LO and TO Raman modes with respect to different FLA energy densities. (b) Raman spectra of FLA treated virgin GaAs, GaAs:In, GaAs:N, GaAs:Zn, under which annealing condition the strongest 1.30 µm PL (for GaAs:Zn the strongest NBE) is observed. The spectra are vertically shifted for better visibility.

Fig. 6 Room temperature PL spectra of FLA treated GaAs:Zn and p-GaAs:N. The 1.30 µm emission is absent from both samples.